\documentclass{emulateapj}
\usepackage{apjfonts}

\shorttitle{Shallow X-Ray Afterglows from Inhomogeneous GRB Jets}
\shortauthors{Toma et al.}
\begin{document}
\title{Shallow Decay of Early X-ray Afterglows from Inhomogeneous
Gamma-Ray Burst Jets
}
\author{Kenji Toma\altaffilmark{1}, Kunihito Ioka\altaffilmark{1}, 
Ryo Yamazaki\altaffilmark{2}, and Takashi Nakamura\altaffilmark{1}}

\altaffiltext{1}{Department of Physics, Kyoto University,
Kyoto 606-8502, Japan; toma@tap.scphys.kyoto-u.ac.jp}
\altaffiltext{2}{Department of Physics, Hiroshima University,
Higashi-Hiroshima 739-8526, Japan}
\begin{abstract}
Almost all the X-ray afterglows of gamma-ray bursts (GRBs) 
observed by the {\it Swift} satellite
have a shallow decay phase in the first thousands of seconds.
We show that in an inhomogeneous jet 
(multiple-subjet or patchy-shell) model 
the superposition of the afterglows of off-axis subjets 
(patchy shells) can have the shallow decay phase.
The necessary condition for obtaining the shallow decay phase is
that $\gamma$-ray bright subjets (patchy shells) should have
$\gamma$-ray efficiency higher than previously estimated,
and should be surrounded by 
$\gamma$-ray dim subjets (patchy shells) with low $\gamma$-ray efficiency.
Our model predicts that events with dim prompt 
emission have the conventional afterglow light curve 
without the shallow decay phase like GRB 050416A.
\end{abstract}


\section{Introduction}
\label{sec:intro}

Before the {\it Swift} era, most of the X-ray and optical afterglows 
of gamma-ray bursts (GRBs) were detected only several hours after 
the burst trigger.
{\it Swift} observations are unveiling the first several 
hours of the afterglows 
\citep[e.g.,][]{taglia05,chinca05,nousek05,cusuma05,hill05,vaughan05}.
Recently,
\citet{nousek05} analyzed the first 27 afterglows detected by 
{\it Swift} XRT, and reported that almost all the early X-ray afterglows of 
{\it Swift} GRBs do not show a simple power-law flux decline.
They show a ``canonical'' behavior, where 
the light curve begins with a very steep decay, 
turns into a very shallow decay $\sim t^{-0.5}$, and 
finally connects to the conventional late-phase afterglow $\sim t^{-1}$
which is similar to what was observed in the pre-{\it Swift} era.

The shallow decay phase implies that more 
time-integrated radiation energy is observed at later
time. 
This is unexpected in the standard model that can explain
the late-phase afterglows, i.e., 
the synchrotron shock model of an impulsive homogeneous jet
\citep[][for reviews]{zhangmes04,piran04}.
There seems to be essentially no spectral variation at the transition
from the shallow decay phase to the conventional decay phase.
This suggests that the origin of the transition is either hydrodynamical
or geometrical. 

In the hydrodynamical model, the GRB jet is 
not impulsive but the energy is injected continuously into the blast wave 
\citep[][and references therein]{zhang05,nousek05,panai05,granot05}.
Such a continuous injection can be realized either 
by the long-lived central engine or
the short-lived central engine  with some distribution of 
the Lorentz factors of the launched shells.
In the case of the long-lived central engine,
the more time-integrated injected energy is required in later time 
while the injection should be stopped abruptly at some time ($\sim 10^4$~s).
In the case of the short-lived central engine,  
slower shells should have more energy than faster ones and
a lower cut-off of  the Lorentz factor should exist.
Since the afterglow is
dim in the shallow decay phase, the $\gamma$-ray efficiency 
for the front shells is much higher 
than previously estimated both 
in the long-lived central engine case and 
in the short-lived central engine case.   
This is  problematic in the framework of the internal shock model.

In the geometrical model, it is assumed that we observe more energetic
regions of the GRB jet later as the afterglow
shock decelerates and the visible region increases. 
The shallow decay phase of the ``canonical'' afterglow may be a 
combination of the tail
part of the prompt emission and the delayed afterglow 
emission from an off-axis jet \citep{eichler05}.
In this picture, the duration and the flatness
of the shallow decay phase correlate with the spectral peak photon energy
$E_p$ and the isotropic $\gamma$-ray energy $E_{\gamma,iso}$, because
all these quantities depend on the viewing angle. 
The jet break occurs just after the off-axis afterglow is observed,
so that  the conventional decay phase $(\sim t^{-1})$ is expected to
be short.  
Since \citet{eichler05} discussed a specific ``ring-shaped'' jet,
more general studies for the jet angular structure
are desirable to know the general characteristics of geometrical model
\citep[see also][]{panai05}.

\begin{figure}
\epsscale{0.80}
\plotone{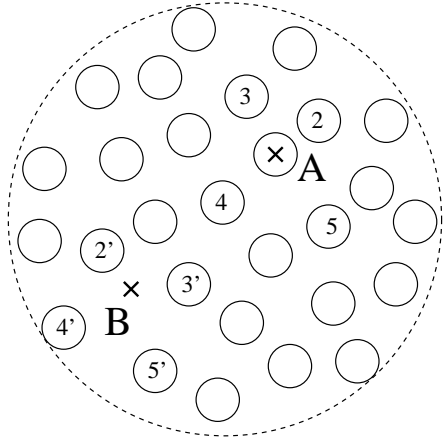}
\caption{
Setup for our analysis of an inhomogeneous jet.
A whole jet ({\it dashed circle}) consists of multiple subjets
({\it solid circles}).
Points `A' and `B' describe the lines of sight for our calculations.
We take the initial opening half-angle
of the subjets and the whole jet as $\Delta\theta_0^i=0.01$~rad
and $\Delta\theta_0^w=0.1$~rad.
Subjets $(2), (3), (4),$ and $(5)$ for the line of sight `A'
(similarly, $(2'), (3'), (4'),$ and $(5')$ for the line of sight `B')
have the viewing angles
$\theta_v^i = 0.025, 0.03, 0.035,$ and $0.04$~rad.
}
\label{fig:agjet}
\end{figure}

In this Letter, we develop an inhomogeneous jet model to reproduce
the ``canonical'' X-ray afterglows of GRBs in the framework of the
geometrical model.
In order to study the angular energy distribution in the jet,
we consider an extremely inhomogeneous jet (a multiple-subjet model).
Figure~\ref{fig:agjet} illustrates the setup for our analysis of
an inhomogeneous jet.
We assume that the whole jet ({\it dashed circle}) consists of multiple 
subjets ({\it solid circles}), and the energy injected among subjets
is negligible compared to the energy inside each subjet.
Each subjet is assumed to make a prompt $\gamma$-ray radiation and a
subsequent afterglow following the standard scenario.
We calculate the early phase of the afterglow by 
superposing the contribution of each subjet, and
study necessary conditions 
for reproducing the ``canonical'' afterglows of the {\it Swift} GRBs.

The inhomogeneous jet models have been used to study the diversity 
of the prompt emission of GRBs \citep{nakamura00,kumar00}.
The geometrical effects in such models can explain the Amati correlation
\citep{toma05,eichler04,yama04} and even the Ghirlanda correlation
\citep{levinson05}.
The patchy-shell model is  also used to explain the observed 
variability of the early afterglow light curve and the polarization of a 
particular event like GRB 021004 \citep[e.g.,][]{nakar04}.

In \S~\ref{sec:model}, we study the necessary conditions for the jet properties
to reproduce the ``canonical'' afterglows.
Summary and discussions are given in 
\S~\ref{sec:discussion}.

%%%%%%%%%%%%%%
%%%%%%%%%%%%

\section{Necessary conditions for the shallow decay of early X-ray 
afterglows}
\label{sec:model}

Figure~\ref{fig:agjet} shows an example of the initial 
jet structure.
We may consider the initial opening half-angle of each subjet 
$\Delta\theta_0^i$ to be $\gtrsim {\Gamma_0^i}^{-1}$, 
where $\Gamma_0^i \simeq 10^2 - 10^3$ is the initial Lorentz factor 
of each subjet.
The superscripts `$i$' and `$w$' denote each subjet and the whole jet, 
respectively, while the subscript `$0$' denotes 
the initial time when each subjet begins to
decelerate.
Each subjet is assumed to emit the prompt emission by the internal shock
and the subsequent afterglow by the synchrotron emission from the 
external shock of an impulsive homogeneous jet.
We assume that all the subjets are ejected at essentially the same time,
i.e., over a period that is much shorter than the  timescale of the afterglow.

In the following, we discuss the necessary conditions for explaining the 
``canonical'' behavior of X-ray afterglows of {\it Swift} GRBs.
The discussion is separated into two cases:
Case $(i)$ the line of sight is along a subjet. 
Case $(ii)$ the line of sight is off-axis for any subjet.
For both cases we will obtain the necessary conditions to reproduce
the ``canonical'' afterglow.

%%%%%%%%%%%%%%
\begin{figure}
\plotone{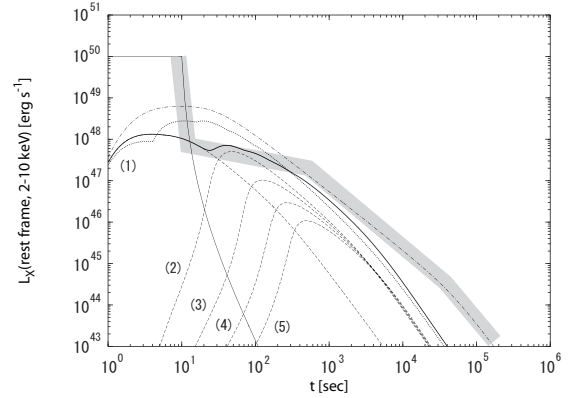}
\caption{
Example of the afterglow light curve 
(the isotropic-equivalent luminosity)
in the range $2 - 10$~keV
measured in the cosmological rest frame of the GRB.
The dot-dashed line is the afterglow from a jet
with $E_{k,iso}^w=10^{52}$~erg, $\Delta\theta_0^w=0.1$~rad,
and $\theta_v=0$.
The thin solid flat line for $t<10$~s
represents a typical prompt burst 
which corresponds to the late-phase of the dot-dashed-line afterglow.
The thin solid curved line $t>10$~s represents the tail part of the 
prompt emission which we set as
$\propto (t-9.0)^{-3.5}$.
The dashed line $(1)$ is the afterglow from a subjet
with $E_{k,iso}^i=3\times10^{51}$~erg, 
$\Delta\theta_0^i=0.01$~rad, and $\theta_v^i=0$.
The dashed lines $(2), (3), (4),$ and $(5)$ are the afterglows
from subjets with $E_{k,iso}^i=3\times10^{52}$~erg,
$\Delta\theta_0^i=0.01$~rad, and $\theta_v^i=0.025, 0.03, 0.035,$
and $0.045$~rad, respectively.
These subjets correspond to $(2), (3), (4),$ and $(5)$ for the 
line of sight `A' (or $(2'), (3'), (4'),$ and $(5')$ for the line 
of sight `B') in Fig~\ref{fig:agjet}.
The thick solid line is the superposition of all the dashed lines
$(1)-(5)$.
The shaded line is what we expect for the afterglows from 
inhomogeneous GRB jets.
For the dotted line, see \S~\ref{sec:discussion}.
}
\label{fig:f3}
\end{figure}
%%%%%%%%%%%%

\paragraph{Case $(i)$}
\label{sec:case1}

In this case the line of sight is, for example, `A' in Fig~\ref{fig:agjet}.
The shaded line in Fig~\ref{fig:f3} shows the afterglow light curve
in the range of $2-10$~keV obtained in our calculation.   
This demonstrates that in case $(i)$ the ``canonical'' afterglow light curve
can be obtained under certain conditions explained below.

We calculate X-ray afterglow emission from the external shock of 
an impulsive homogeneous jet with sharp edge following
\citet{panai01}.
The jet dynamics is calculated by the mass and energy conservation
equations with the effect of sideways expansion at the
local sound speed and radiative energy losses.
The initial radius of the shell is set to be $0.01$ times the 
deceleration radius.
For the calculation of the synchrotron emission, the spectrum is 
approximated as a piecewise power law with the injection break $\nu_m$
and the cooling break $\nu_c$.
We neglect the self absorption 
break because we focus on the spectrum for 
$\nu > {\rm min}(\nu_m, \nu_c)$.
The received flux is calculated by integrating over the equal arrival
time surface of photons to the observer.
Neither the synchrotron self-Compton emission
nor the reverse shock emission is taken into account, for simplicity.
In all the following calculations, we fix  
the initial Lorentz factor of the shell as $\Gamma_0 = 300$, 
the initial opening half-angle of the subjet as 
$\Delta\theta_0^i=0.01$~rad,
the number density of the circumburst medium as $n=1~{\rm cm}^{-3}$,
the ratio of the magnetic energy and the 
accelerated electron energy to the shocked thermal energy 
as $\epsilon_B=0.01$ and $\epsilon_e=0.1$, 
respectively, 
and the index of the energy distribution function of the accelerated 
electrons as $p=2.3$.

In Fig~\ref{fig:f3}, the dot-dashed line represents 
the afterglow light curve expected before the {\it Swift} era,
i.e., the afterglow from a homogeneous jet 
with a typical afterglow energy of $E_{k,iso}^w=10^{52}$~erg,
an opening half-angle of $\Delta\theta^w_0=0.1$~rad,
and a viewing angle of $\theta_v=0$.
The X-ray afterglow emission has a rising light curve peaking at 
the shell deceleration time $t_{\rm dec} \simeq 
5~(E_{k,iso}^w/10^{52}~{\rm erg})^{1/3}
(\Gamma_0/300)^{-8/3} n^{-1/3}$~s. 
Around this time the XRT band is crossed by $\nu_m$ and $\nu_c$ for typical
parameters \citep{sari98}, so that after the peak,
the light curve shows a smooth decline of $\sim t^{-1.2}$.
The jet break time
is estimated as $t_{\rm jet}^w \simeq 
2\times10^4~{\rm s}~
(\Delta\theta_0^w/0.1~{\rm rad})^{8/3}(E_{k,iso}^w/
10^{52}~{\rm erg})^{1/3} n^{-1/3}$~s \citep{sari99}.
After this time the light curve steepens into $\sim t^{-2.3}$,
although the steepening is gradual 
\citep[see][]{kumarpanai00}.

The thin solid flat line around $L_X \sim 10^{50}~{\rm erg}~{\rm s}^{-1}$
and for $t<10$~s
represents a typical prompt burst
with a duration of $\simeq 10$~s.
The isotropic X-ray energy is about $\sim 10^{51}$ erg,
and for the typical GRB spectrum 
$\nu F_{\nu} \propto \nu$ at low energy,
the isotropic $\gamma$-ray energy should be about
$\gtrsim 10^{52}$~erg.
This is comparable to or larger than 
the afterglow energy $E_{k,iso}^w=10^{52}$~erg
as in the actual observations \citep{lloyd04}.
The thin curved solid line for $t>10$~s is the tail part of the prompt
burst, which comes from the region with large viewing angles in the 
whole jet.
The temporal index of the tail part can be approximated 
as $\sim -1+\beta$, where $\beta\sim -2.5$ is 
the high energy photon index of the prompt
emission \citep[e.g.,][]{zhang05}.
Even if the emission regions are so patchy, the tail emission
may be smooth since pulses from large viewing 
angles have long duration and overlap with each other \citep{yama05}.

First, consider  
the on-axis subjet which includes the line of sight `A' in 
Fig~\ref{fig:agjet}.
If the afterglow energy of the on-axis subjet is as large as
$E_{k,iso}^i = 10^{52}$~erg, the afterglow flux is comparable to
that of the dot-dashed line. 
Then it overwhelms
the tail part of the prompt emission, and the temporal index of 
the afterglow emission just after the prompt burst will be 
$\sim t^{-1.2}$ or $\sim t^{-2.3}$, which is inconsistent with 
the steep decay observed by XRT.
The dashed line $(1)$ in Fig~\ref{fig:f3} is the afterglow 
emission from the on-axis subjet with $E_{k,iso}^i=3 \times 10^{51}$~erg. 
Compared to the dot-dashed-line afterglow with 
$E_{k,iso}^w=10^{52}$~erg, we see that the deceleration time $t_{\rm dec}$ is 
a little earlier, and the peak luminosity is smaller since the 
spectral peak flux is $F_{\nu,{\rm max}} \propto E_{k,iso}$.
The jet break time of the subjet is much smaller because of 
strong dependence of $t_{\rm jet}^i$ on $\Delta\theta_0^i$, and 
is estimated as $t_{\rm jet}^i \simeq 
30~(\Delta\theta_0^i/0.01~{\rm rad})^{8/3}(E_{k,iso}^i/
3\times10^{51}~{\rm erg})^{1/3} n^{-1/3}$~s.
In this case the steep decay due to the tail part of the prompt emission
can be observed.
Therefore the afterglow energy $E_{k,iso}^i$ of the on-axis subjet
should be 
at most $1/3$ of that of the dot-dashed line which is
typical before the {\it Swift} era.

Secondly, we can show that the shallow afterglow can be 
produced by the superposition of the subjet emissions.
In Fig~\ref{fig:f3} we show the afterglow emissions from the 
off-axis subjets, which do not include the line of sight `A'.
The dashed lines $(2), (3), (4),$ and $(5)$ are 
the afterglow emissions from the subjets with $\theta_v^i=0.025,~
0.03,~0.035,$ and $0.04$~rad, respectively.
These subjets are illustrated in Fig~\ref{fig:agjet}
and have equal afterglow energies 
$E_{k,iso}^i=3\times10^{52}$~erg.
This is larger than that of the dot-dashed line by a factor 3.
The time at the peak is when the emission from the edge of 
the subjet arrives at the observer, and is
larger for the subjet with larger $\theta_v^i$ \citep{granot02}.
The superposed light curve of the on-axis and 
off-axis subjets is 
described by the thick solid line, which shows a shallow decline
compared to the conventional decline $\sim t^{-1.2}$.
If all the off-axis subjets have equal viewing
angles, the superposition of their contributions produce a bump
in our calculation.
Nevertheless, the real afterglow may be flat
because 2D hydrodynamical simulations
show that a rising part of the light curve
when viewed with 
$\Delta\theta_0^i \lesssim \theta_v^i \lesssim 2\Delta\theta_0^i$
is much flatter than 1D calculations like ours
\citep{granot02}.

All the subjets expand sideways, and then begin to merge with
each other.
They will cease to expand sideways because of their pressure, and
finally merge into one shell producing the conventional afterglow
emission.
Although we cannot follow the merging process by our simple calculations,
the merged whole jet would make the conventional
decline of the dot-dashed line at the late time, 
since the $E_{k,iso}^i$ averaged over the solid angle
is similar to $E_{k,iso}^w = 10^{52}$~erg.
Therefore we suppose that the shallow decay phase 
would smoothly connect to the dot-dashed line 
and the final afterglow would be like the shaded line.

The prompt emission is dominated by that from the on-axis subjet
because of the beaming effect.
Thus the prompt burst energy $E_{\gamma,iso}^i$ of the on-axis 
subjet is $\gtrsim 10^{52}$~erg.
Since $E_{k,iso}^i \sim 3 \times 10^{51}$ erg,
this implies that the $\gamma$-ray efficiency for the on-axis subjet
is $\epsilon_{\gamma} \equiv 
E_{\gamma,iso}^i/(E_{\gamma,iso}^i + E_{k,iso}^i) \gtrsim 75\%$, 
which is larger than previously estimated.
This requirement is similar to the hydrodynamical models for
the shallow decay afterglows.

Now what is observed when our line of sight is along the subjet with
an energetic afterglow of $E_{k,iso}^i=3\times10^{52}$~erg?
Let us assume that the ``canonical'' afterglow is observed also
in this case.
Then the energy of the prompt emission should be 
$E_{\gamma,iso} \gtrsim 10^{53}$~erg in order for the tail emission
to be larger than  the afterglow emission from the on-axis subjet.
From the necessary condition for the shallow decay phase 
obtained in the above discussion, the number of the energetic 
afterglow subjets should be larger than that of 
the high $\gamma$-ray efficiency subjets.
This leads to larger event rate of more energetic prompt bursts, 
which is not consistent with current observations.
Therefore the subjets with energetic afterglows should have low
$\gamma$-ray efficiency and dim prompt emissions so that
they are hard to be observed. 

In summary, 
a subjet making a bright prompt burst should have
a dim afterglow 
and should be surrounded by several
subjets with dim prompt bursts and bright afterglows.
A favorable GRB jets may have discrete spots with bright bursts
and dim afterglows surrounded by the regions 
with dim bursts and bright afterglows.

\paragraph{Case $(ii)$}
\label{sec:case2}

Next, we consider the necessary condition  under which the ``canonical'' 
afterglow is
observed when our line of sight is off-axis for any subjet like `B' in 
Fig~\ref{fig:agjet}.
The ``canonical'' afterglow light curve is obtained by the 
same calculation as in  case $(i)$ removing the contribution from 
the on-axis subjet.
The afterglow light curves from the subjets $(2'), (3'), (4'),$ and $(5')$
in Fig~\ref{fig:agjet}
are the same as the dashed lines $(2), (3), (4),$ and $(5)$, respectively.
The nearest subjet $(2')$ should have a viewing angle of
$\theta_{v,{\rm min}} \sim 2\Delta\theta_0^i$,
because if $\theta_v^i < \theta_{v,{\rm min}}$
the contribution of the afterglow 
emission overwhelms the tail part of the prompt emission
while if $\theta_v^i > \theta_{v,{\rm min}}$
a rising afterglow appears after the tail
of the prompt emission.
The predicted total afterglow light curve in this case $(ii)$ is similar
to that in case $(i)$, i.e., the shaded line.

In case $(ii)$ also,
the $\gamma$-ray efficiency $\epsilon_{\gamma}$ 
should be large.
The prompt emission is dominated by the subjets with
the viewing angles $\theta_v^i \sim \theta_{v,{\rm min}}$.
If the velocity of a point source has an angle 
$\theta$ with the line of sight,
the observed energy from this source is
$\propto (1-\beta\cos\theta)^{-3}$ because of the beaming effect.
The observed energy from widely distributed segments of size 
$\Delta\theta_0^i$ with similar viewing angles 
$\theta_{v,{\rm min}}$ roughly follows 
$E_{\gamma,iso} \propto [1-\beta_0^i\cos(\theta_{v,{\rm min}}-
\Delta\theta_0^i)]^{-2}$ \citep{toma05,eichler04,levinson05}.
This is derived by the integration of the contribution of the point
source over solid angle occupied by the emission regions.
Thus, in this case with
$\theta_{v,{\rm min}} \sim 2\Delta\theta_0^i$, we receive 
the prompt burst energy 
$E_{\gamma,iso} \sim (1-\beta_0^i\cos\Delta\theta_0^i)^{-2} 
E_{\gamma,iso}^i \simeq (\Gamma_0^i\Delta\theta_0^i)^{-4} 
E_{\gamma,iso}^i$, where $E_{\gamma,iso}^i$ is the isotropic 
energy of the prompt emission when a subjet is viewed on-axis.
The received prompt energy is 
$E_{\gamma,iso} \gtrsim 10^{52}$~erg in the above calculation,
and thus $E_{\gamma,iso}^i \gtrsim 10^{54}$~erg.
Since $E_{k,iso}^i = 3\times10^{52}$~erg, we obtain
$\epsilon_{\gamma} \gtrsim 97\%$.
If this case occurs dominantly over the case $(i)$, 
we should observe many very bright GRBs,
when the line of sight is along the $\gamma$-ray bright subjet.
Thus the contribution of this case to the shallow decay afterglows
would be small. 

\clearpage

\section{Discussion}
\label{sec:discussion}

We have investigated early X-ray afterglows of GRBs within 
inhomogeneous jet models by using a multiple-subjet model.
We find that several off-axis subjets can reproduce the 
shallow decay phase of the light curves observed by 
{\it Swift} XRT.
The shallow decay phase is produced by the superposition of 
the afterglows from off-axis subjets, and it connects to
the conventional late-phase afterglow produced by the merged whole jet.

We claim that the shallow decay phase arises prior to the merging
of the subjets.
So the shape of the early afterglow light curve depends sensitively
on the assumed sideways expansion speed of the subjets.
The sideways expansion speed of the jet is highly uncertain and
has been debated by using hydrodynamical calculations 
\citep[e.g.,][]{kumar03,cannizzo04}.
In this Letter, we have assumed that each subjet expands sideways
at the local sound speed.
In Fig~\ref{fig:f3} we also show an afterglow light curve
(a superposition of the $(1)-(5)$ fluxes) calculated under the assumption 
that each subjet expands at the local light speed 
({\it dotted line}).
The fluxes from the off-axis subjets peak earlier, so that 
the shallow decay phase disappears.
If the local expansion speed is varied from the sound one to the light 
one, light curves varies from the thick solid line to the dotted one
smoothly.
For the local light speed case,
we can obtain the shallow decay phase if the subjets
are distributed more sparsely, for example, with
$\theta_v^i = 0.032, 0.037, 0.042,$ and $0.047$~rad
for the off-axis subjets.
However, if each subjet makes a hot cocoon which expands 
relativistically in the lab frame, all the subjets would merge 
around the deceleration time
and the light curve would be the conventional one,
i.e., dot-dashed line in Fig~\ref{fig:f3}.

We found  necessary conditions for obtaining the 
``canonical'' afterglow by
separating our discussions into two cases, i.e.,
whether the line of sight is along a subjet (case $(i)$)
or not (case $(ii)$).
In both cases $(i)$ and $(ii)$, subjets producing a bright prompt emission should
have $\gamma$-ray efficiency larger than previously estimated.
This requirement is similar to the hydrodynamical model
\citep{zhang05,nousek05} and
is problematic in the framework of the internal shock model.

There are some predictions in our model.
First, in case $(i)$, a subjet producing a bright prompt burst
should have a dim afterglow emission and should be surrounded by 
several subjets producing a dim prompt and a bright afterglow
emissions.
The possibility of such a jet structure cannot be excluded
at present and should be tested by future observations.
When the line of sight is along the subjet with a dim prompt and a 
bright afterglow emissions, the conventional
afterglow light curve without a shallow phase is observed.
Therefore we predict that small $E_{\gamma,iso}$ events should have the 
conventional afterglow light curve.
Among 10 {\it Swift} GRBs with known redshifts, GRB 050416A
has an extremely small $E_{\gamma,iso}$ of $\lesssim 10^{51}$~erg
and does not have a shallow decay phase \citep{nousek05}.
This event may support the case $(i)$ of the inhomogeneous jet model,
although more statistics is required to confirm the validity.
Secondly, the number of subjets with dim bursts and bright afterglows
should be several times larger than 
that of the observed $\gamma$-ray bright subjet.
Then the true GRB rate should be several times larger than the current
estimates.
In addition, since many subjets are $\gamma$-ray dark,
the mean $\gamma$-ray efficiency over the whole jet does not need 
to be so large \citep{kumar00}.
Only a subjet that happens to emit almost all energy into $\gamma$-ray may 
be observed as a GRB.

Case $(ii)$ suggests that for most events both the prompt and 
the afterglow emissions arise from off-axis viewing angles,
which is similar to the scenario of \citet{eichler05}.
In this case, we found that most of the subjets should produce 
a bright prompt and a dim afterglow emissions.
When the line of sight is along such a subjet, the conventional
but dim afterglow is observed.
Then we predict that there should be large $E_{\gamma,iso}$ events
with a conventional afterglow in case $(ii)$.
We should observe such bright $\gamma$-ray events with a similar order of rate as 
the ``canonical'' events, which may be tested in future.

\acknowledgements
We thank G.~Sato and T.~Takahashi for useful discussions.
This work is supported in part by
Grant-in-Aid for the 21st Century COE
``Center for Diversity and Universality in Physics''
from the Ministry of Education, Culture, Sports, Science and Technology
(MEXT) of Japan
and also by Grants-in-Aid for Scientific Research
of the Japanese Ministry of Education, Culture, Sports, Science,
and Technology 14047212 (K.~I. and T.~N.), 14204024 (K.~I. and T.~N.)
and 17340075 (T.~N.).

%\clearpage

\end{document}